# Electromagnetic Potentials Basis for Energy Density and Power Flux

Harold E. Puthoff
Institute for Advanced Studies at Austin, 11855 Research Blvd., Austin, Texas 78759

E-mail: puthoff@earthtech.org
Tel: 512-346-9947
Fax:512-346-3017

**Abstract** It is well understood that various alternatives are available within EM theory for the definitions of energy density, momentum transfer, EM stress-energy tensor, and so forth. Although the various options are all compatible with the basic equations of electrodynamics (e.g., Maxwell's equations, Lorentz force law, gauge invariance), nonetheless certain alternative formulations lend themselves to being seen as preferable to others with regard to the transparency of their application to physical problems of interest. Here we argue for the transparency of an energy density/power flux option based on the EM potentials alone.

## 1. Introduction

The standard definition encountered in textbooks (and in mainstream use) for energy density $u$ and power flux $\mathbf{S}$ in EM (electromagnetic) fields is given by

$$u(\mathbf{r},t)=\frac{1}{2}\left[\varepsilon_0|\mathbf{E}|^2+\mu_0|\mathbf{H}|^2\right] \quad (1.a), \quad \mathbf{S}(\mathbf{r},t)=(\mathbf{E}\times\mathbf{H}) \quad (1.b).$$

One can argue that this formulation, even though resulting in paradoxes, owes its staying power more to historical development than to transparency of application in many cases. One oft-noted paradox in the literature, for example, is the (mathematical) apparency of unobservable momentum transfer at a given location in static superposed electric and magnetic fields, seemingly implied by (1.*b*) above.[1] A second example is highlighted in Feynman's commentary that use of the standard $(\mathbf{E},\mathbf{H})$ Poynting vector approach leads to "… a peculiar thing: when we are slowly charging a capacitor, the energy is not coming down the wires; it is coming in through the edges of the gap," a seemingly absurd result in his opinion, regardless of its uniform acceptance as a correct description [1].

As developed in detail here, definitions for energy transfer based on the use of EM potentials alone is shown to yield results more in keeping with intuition for conundrums such as the above, and in the process provide additional insight into the significance of the potentials (even in classical theory) that extend beyond simply that of constituting a placeholder for EM fields.

[1] This has on occasion led to time-consuming debates in the engineering literature as to the feasibility of certain forms of electromagnetic propulsion.

## 2. Background

Though little discussed, there exists a rich literature on alternative definitions of EM energy density and power flux, the general purpose of which has been to attempt to bring clarity to some of the ambiguities associated with the standard definitions. In an early paper by Slepian, for example, no less than eight alternatives to the standard definitions are considered, all congruent with Maxwell's equations [2]. That is, once integrated over a volume of interest the net energy and power flow resulting from use of the various alternative expressions are all in agreement; only the density *distributions* and their associated interpretations differ.

As an example, for Maxwell fields the standard expression involving energy density and power flux

$$\frac{\partial}{\partial t}u + \nabla\cdot\mathbf{S} = \frac{\partial}{\partial t}\frac{1}{2}\left(\varepsilon_0\left|\mathbf{E}\right|^2 + \mu_0\left|\mathbf{H}\right|^2\right) + \nabla\cdot\left(\mathbf{E}\times\mathbf{H}\right) \tag{2}$$

is trivially found by the use of vector identities to be equally satisfied by

$$u = \frac{1}{2}\left(\varepsilon_0\left|\mathbf{E}\right|^2 + \mu_0\left|\mathbf{H}\right|^2\right) - \nabla\cdot\mathbf{Y} \quad (3.a), \quad \mathbf{S} = \mathbf{E}\times\mathbf{H} + \nabla\times\mathbf{X} + \frac{\partial}{\partial t}\mathbf{Y} \quad (3.b)$$

where $\mathbf{X}$ and $\mathbf{Y}$ are arbitrary vector fields [3].

Application of the above option for the case $\mathbf{X} = \varphi\mathbf{H}$, $\mathbf{Y} = \mathbf{0}$, yields [4]

$$\mathbf{S} = -\frac{\partial\mathbf{A}}{\partial t}\times\mathbf{H} + \varphi\left(\mathbf{j} + \varepsilon_0\frac{\partial\mathbf{E}}{\partial t}\right), \tag{3}$$

where $(\mathbf{A},\varphi)$ are the vector and scalar potentials defined by

$$\mathbf{E} = -\nabla\varphi - \frac{\partial\mathbf{A}}{\partial t}, \quad \mathbf{B} = \nabla\times\mathbf{A}. \tag{4}$$

From (3) we find that this alternative form for the power flux $\mathbf{S}$ addresses, and provides alternative interpretations for, the conundrums referred to above in that (a) by virtue of the time derivatives no energy flow is attributed to static superposed electric and magnetic fields in free space, and (b) given that $\mathbf{S}\propto\varphi\mathbf{j}$ detailed calculation (see Section 4.2) shows that energy can be seen to flow down the wire during the charging of a capacitor. Therefore, at least for this pair of problems, the definition of $\mathbf{S}$ given by (3) appears more aligned with our intuition than (1.b) relative to the physical processes involved.

Unfortunately, in the absence of a well-structured theoretical guideline, selection of the above and similar modifications, typically involving mixed field and potential terms, remains an *ad hoc* procedure at best. In the next Section we consider a universally-applicable, systematic treatment based on the use of EM potentials alone.

**3. Potentials-based Definitions**

Although use of the vector and scalar potentials $(\mathbf{A},\varphi)$ in place of the EM fields $(\mathbf{E},\mathbf{B})$ is considered simply to be an option in classical theory, in quantum theory they are understood to be more fundamental than the derivative electric and magnetic fields $(\mathbf{E},\mathbf{B})$ which are the "coin of the realm" in ordinary classical theory. In classical electrodynamics the choice of which variable pair to use is arbitrary, and the overall resulting predictions in terms of observables are indistinguishable. Nonetheless, cogent arguments can be made that the $(\mathbf{A},\varphi)$ approach is to be preferred in many cases, even in classical EM theory, because of increased transparency in application.

By virtue of the freedom in EM theory to choose a gauge (gauge invariance), when employing the $(\mathbf{A},\varphi)$ potentials it is convenient to choose the Lorentz gauge[2]

$$\nabla\cdot\mathbf{A}=-\frac{1}{c^2}\frac{\partial\varphi}{\partial t}. \tag{5}$$

This results in simplified wave equations fully equivalent to Maxwell's equations for **E** and **B,**

$$\nabla^2\varphi-\frac{1}{c^2}\frac{\partial^2\varphi}{\partial t^2}=-\frac{\rho}{\varepsilon_0},\quad \nabla^2\mathbf{A}-\frac{1}{c^2}\frac{\partial^2\mathbf{A}}{\partial t^2}=-\mu_0\mathbf{j}, \tag{6}$$

in which the scalar potential $\varphi$ is determined by the charge density $\rho$ alone, and the vector potential $\mathbf{A}$ is determined by the current density $\mathbf{j}$ alone. Key to the development here, it is the dependence on separate source terms for the two potentials that contributes to an independence that leads to transparency in application (see below). The solutions to Eqns. (6) are given by the retarded Green's functions

$$\varphi(\mathbf{r},t)=\int\frac{\rho\left(\mathbf{r}',t-|\mathbf{r}-\mathbf{r}'|/c\right)}{4\pi\varepsilon_0|\mathbf{r}-\mathbf{r}'|}dV',\quad \mathbf{A}(\mathbf{r},t)=\int\frac{\mu_0\mathbf{j}\left(\mathbf{r}',t-|\mathbf{r}-\mathbf{r}'|/c\right)}{4\pi|\mathbf{r}-\mathbf{r}'|}dV'. \tag{7}$$

By use of these equations and the definitions provided in (4) the usual Maxwell equations in terms of **E** and **B** as driven by charge and current densities can be rederived.

[2] We note in passing that should evidence ever be forthcoming for a finite photon mass, regardless of how infinitesimal, then gauge invariance would be broken and the Lorentz gauge would be required [5].

It is at this juncture that our approach differs substantially from the usual approach concerning the definitions of EM energy density and power flux, and that as a consequence can provide for increased transparency in application. First, we note that the scalar potential $\varphi$ and vector potential $\mathbf{A}$ each satisfy independent wave equations. For this case in the theory of partial differential equations there arise, specifically in terms of gradients, natural measures of associated 'energy density' and 'power flux' (and associated conservation laws) for the dependent variables of interest (here $\varphi$ and $\mathbf{A}$) [6]. Secondly, we note from (6) that the field effects from charges and currents can be seen as resulting from two *noninteracting* physical processes described, respectively, by the scalar and vector potentials $\varphi$ and $\mathbf{A}$. We can therefore anticipate that reasonable and satisfactory definitions for the densities of momenta and energy should be expressible in terms of *unmixed* derivatives of $\varphi$ and $\mathbf{A}$. Though little referenced, such approaches have been advanced in the literature from time to time in various forms and constitute the basis for the present investigation [7,8].

In place of the standard definition for EM energy density, (1.a), we take as our definition

$$u(\mathbf{r},t)=u_A-u_\varphi+\rho\varphi, \tag{8}$$

where $u_A$ is an energy density defined in terms of derivatives of the vector potential only,[3]

$$u_A(\mathbf{r},t)=\frac{1}{2\mu_0}\sum_i\left[\frac{1}{c^2}\left(\frac{\partial A_i}{\partial t}\right)^2+\left|\nabla A_i\right|^2\right], \tag{9}$$

and $u_\varphi$ is an energy density defined in terms of derivatives of the scalar potential only,

$$u_\varphi(\mathbf{r},t)=\frac{1}{2}\varepsilon_0\left[\frac{1}{c^2}\left(\frac{\partial\varphi}{\partial t}\right)^2+\left|\nabla\varphi\right|^2\right]. \tag{10}$$

In place of the standard definition for EM power flux (1.b), we take as our definition

$$\mathbf{S}(\mathbf{r},t)=\mathbf{S}_A-\mathbf{S}_\varphi+\varphi\mathbf{j}, \tag{11}$$

with $\mathbf{S}_A$ and $\mathbf{S}_\varphi$ defined in terms of their respective (and separate) derivatives of the potentials as well,

---

[3] For simplicity of notation it is to be understood here that in the application of Equations (9) and (12) the gradient operator $\nabla=\hat{\mathbf{1}}_x\frac{\partial}{\partial x}+\hat{\mathbf{1}}_y\frac{\partial}{\partial y}+\hat{\mathbf{1}}_z\frac{\partial}{\partial z}$ and vector components $A_i$ are to be expressed in terms of rectangular coordinates $i=x,y,z$.

$$\mathbf{S}_A(\mathbf{r},t) = -\frac{1}{\mu_0}\sum_i \left(\frac{\partial A_i}{\partial t}\right)\nabla A_i \,, \tag{12}$$

$$\mathbf{S}_\varphi(\mathbf{r},t) = -\varepsilon_0\left(\frac{\partial \varphi}{\partial t}\right)\nabla \varphi \,. \tag{13}$$

The associated Lorentz power density is given by an expression that parallels that based on densities defined in terms of the electric and magnetic fields $(\mathbf{E},\mathbf{B})$,

$$p_L = -\frac{\partial u}{\partial t} - \nabla \cdot \mathbf{S} \,. \tag{14}$$

Finally, it can be shown that the wave equations (6) are the associated Euler-Lagrange variational equations that can be derived from a Lagrangian density that *has no cross-interaction terms* between the scalar and vector potential terms $\varphi$ and $\mathbf{A}$:

$$\mathsf{L} = -\frac{1}{2\mu_0}\left(\partial^\alpha A_\beta\right)\left(\partial_\alpha A^\beta\right) - j^\alpha A_\alpha = \mathsf{L}_A(\mathbf{r},t) - \mathsf{L}_\varphi(\mathbf{r},t) + \mathbf{j}\cdot\mathbf{A} - \rho\varphi \,, \tag{15}$$

where

$$\mathsf{L}_A(\mathbf{r},t) = \frac{1}{2\mu_0}\sum_i\left[\frac{1}{c^2}\left(\frac{\partial A_i}{\partial t}\right)^2 - \left|\nabla A_i\right|^2\right] \tag{16}$$

and

$$\mathsf{L}_\varphi(\mathbf{r},t) = \frac{1}{2}\varepsilon_0\left[\frac{1}{c^2}\left(\frac{\partial \varphi}{\partial t}\right)^2 - \left|\nabla \varphi\right|^2\right]. \tag{17}$$

## 4. Applications

### 4.1 Radiation Fields

For our first example we examine the case of fields radiated from a short dipole antenna (length $l$ small compared with the wavelength of the radiated signal) that carries a current $i = I\sin\omega t$. In this specific case, unlike those that follow in later sections, both the energy density and power flux distributions are found to be of the same form whether evaluated using the standard EM-fields approach or the potentials-based approach being examined in detail here.

For a dipole oriented vertically in the $x$ direction the vector potential at a distance $r >> l$ between dipole and field point is determined from (7) as

$$A_x = \frac{1}{4\pi}\int_{-l/2}^{l/2} \frac{\mu_0 I \sin(\omega t - kr)}{r} dx \approx \frac{\mu_0 Il}{4\pi r}\sin(\omega t - kr), \tag{18}$$

where $k = \omega/c$ - and for dipole length $l$ small compared to wavelength, and $r >> l$, the numerator and denominator of the integrand, respectively, are each nearly constant. In spherical coordinates $\left(A_r = A_x \cos\theta,\ A_\theta = -A_x \sin\theta,\ A_\phi = 0\right)$ we then have

$$A_r = \frac{\mu_0 Il}{4\pi r}\cos\theta \sin(\omega t - kr), \quad A_\theta = -\frac{\mu_0 Il}{4\pi r}\sin\theta \sin(\omega t - kr), \quad A_\phi = 0, \tag{19}$$

where $\theta$ denotes the angle measured from the $x$ axis of the dipole orientation to that of the $r$-directed vector. The corresponding far-field scalar potential is found from (5) to be

$$\varphi = \frac{Il\cos\theta}{4\pi r}\sqrt{\frac{\mu_o}{\varepsilon_0}}\sin(\omega t - kr) \ . \tag{20}$$

The derivative electric and magnetic fields are then found by (4) to be

$$E_r = E_\phi = 0,\ E_\theta = \frac{\mu_0 Il\omega}{4\pi r}\sin\theta\cos(\omega t - kr), \tag{21}$$

$$B_r = B_\theta = 0,\ B_\phi = \frac{\mu_0 Ilk}{4\pi r}\sin\theta\cos(\omega t - kr), \tag{22}$$

where for the far field only terms in $1/r$ are kept.

To determine the energy density $u$ and power flux $\mathbf{S}$ distributions, we apply (1) for the standard EM-field-based approach, and (8) – (13) for the potentials-based approach. (Recall that for the latter, the gradient operator $\nabla$ and vector components $A_i$ are to be expressed in rectangular coordinates.) The result is that for both the standard EM and the potentials-based approaches the outcome is

$$u(\mathbf{E},\mathbf{B}) = u(\mathbf{A},\varphi) = \mu_0\left[\frac{Ilk\sin\theta\cos(\omega t - kr)}{4\pi r}\right]^2, \tag{23}$$

$$\mathbf{S}(\mathbf{E},\mathbf{B}) = \mathbf{S}(\mathbf{A},\varphi) = \mathbf{1}_r\sqrt{\frac{\mu_0}{\varepsilon_0}}\left[\frac{Ilk\sin\theta\cos(\omega t - kr)}{4\pi r}\right]^2 \ . \tag{24}$$

Therefore, in this (as it turns out, atypical) case, both the energy density and power flux distributions are found to be of the same form whether evaluated on the basis of the standard EM-fields approach or the potentials-based approach.

4.2 Superposed Static Electric and Magnetic Fields

For the second example we examine a configuration in which the difference between the standard EM-fields approach and the potentials-based approach is significant with regard to interpretations. Specifically, we re-address the issue of energy transfer for the case of superposed static electric and magnetic fields initially considered in Section 2 above using an alternate route to derive (3). Eqns. (11) - (13) show that with the present potentials-based approach the power flux $\mathbf{S}$ (and associated momentum transfer) depend on time derivatives $\left(\partial A_i/\partial t\right)$and $\left(\partial\varphi/\partial t\right)$ and therefore, again, do not attribute momentum transfer to superposed static field distributions. This is in contrast to the definition of momentum transfer in the standard formulation where power flux (at a point) is defined in terms of a crossed-field Poynting vector product $\mathbf{S}=\mathbf{E}\times\mathbf{H}$. As alluded to earlier, the Poynting vector definition leads to a possible (mistaken) inference that momentum transfer accompanying power flux can be associated with crossed static electric and magnetic fields, even though there are no observable consequences of such (and, worse, the drawing of faulty conclusions that such momentum transfer can lead to, say, propulsive mechanisms). Though once fully integrated over boundary surfaces the two approaches, the potentials-based approach and the fields-based approach, lead to identical results, it is the point-by-point *distributions* that differ, with the $\left(\mathbf{A},\varphi\right)$ approach being more in harmony with our ordinary intuitions concerning the relationship between causal charge/current sources and field effects.

4.3 Power Dissipation in a Conductor

As a third example we consider a segment of a current-carrying wire of radius $R$ and length $L$ oriented in, say, the z direction in which a steady constant-current-density $\mathbf{j}=\sigma\,\mathbf{E}$ flows, also in the z direction, where $\sigma$ is the conductivity. The current-encircling azimuthal magnetic field is found from $\nabla\times\mathbf{H}=\mathbf{j}\ \Rightarrow\ \int\mathbf{H}\cdot\mathbf{dl}=I$ to be $\mathbf{H}=\hat{\mathbf{1}}_\theta\left|\mathbf{j}\right|R/2$. By use of the standard Poynting flux expression (1.b) this leads to an *inwardly-directed* radial power flux and associated net power dissipation

$$\mathbf{S}=\mathbf{E}\times\mathbf{H}=-\hat{\mathbf{1}}_r\frac{\left|\mathbf{j}\right|^2R}{2\sigma}\quad(25.a),\qquad P=\int\mathbf{S}\cdot\mathbf{da}=-\frac{\pi\left|\mathbf{j}\right|^2R^2L}{\sigma}\quad(25.b).$$

Application of the potential-based flux vector given by (11) – (13), on the other hand, gives a *z-directed* power flux aligned with the current flow, viz.,

$$\mathbf{S}=\varphi\mathbf{j}\quad\Rightarrow\quad\mathbf{S}_u=\varphi_u\mathbf{j},\quad\mathbf{S}_l=\varphi_l\mathbf{j},\quad\mathbf{E}=\hat{\mathbf{1}}_z\frac{\varphi_l-\varphi_u}{L}=\frac{\mathbf{j}}{\sigma},\qquad(26)$$

where subscripts $u$ and $l$ refer to upper and lower parts of the wire segment, and the z-directed electric field is expressed in terms of the potential gradient. From the above we calculate the *longitudinally-directed* power flux and associated net power dissipation as

$$P = \int \mathbf{S} \cdot \mathbf{da} = \left(\varphi_u - \varphi_l\right)\left|\mathbf{j}\right|\pi R^2 = -\left|\mathbf{E}\right| L \left|\mathbf{j}\right| \pi R^2 = -\frac{\pi \left|\mathbf{j}\right|^2 R^2 L}{\sigma} . \quad (27)$$

Though leading to the same result as (25.b) with regard to net power dissipation, between the two options the potentials-based expression for the power flux *distribution* (being aligned with and delivered by the current rather than being inwardly field-directed) is, again, more in alignment with our natural intuitions.

Let us now inquire as to the form of the energy density distributions for this example under the two approaches. (It turns out that, unlike the power flux distributions, they are identical – not always the case as will be seen in the example of Section 4.4).

The energy density distribution for a current-carrying wire can be expressed in terms of either the magnetic field ***B*** or the vector potential ***A*** in accordance with the definitions provided in (1.a) and (9), respectively. The two alternatives are given by:

$$u_B\left(\mathbf{r},t\right) = \frac{1}{2}\mu_0 \left|\mathbf{H}\right|^2 = \frac{1}{2\mu_0}\left|\mathbf{B}\right|^2 \quad (28.a), \qquad u_A\left(\mathbf{r},t\right) = \frac{1}{2\mu_0}\sum_i \left|\nabla A_i\right|^2 \quad (28.b),$$

where, again, it is understood that $\left|\nabla A_i\right|^2$ is to be expressed in Cartesian coordinates (x,y,z) for calculational purposes.

In a constant-current-density $\mathbf{j}$, current-carrying wire oriented in the z direction, the $\mathbf{A}$ and $\mathbf{B}$ field components in cylindrical coordinates, related by $\mathbf{B} = \nabla \times \mathbf{A}$, are given by [9]

$$B_r = B_z = 0, \quad B_\theta = \frac{\mu_0 \left|\mathbf{j}\right| r}{2}, \quad A_r = A_\theta = 0, \quad A_z = -\frac{\mu_0 \left|\mathbf{j}\right| r^2}{4}, \qquad r \le R \quad (29.a)$$

$$B_r = B_z = 0, \quad B_\theta = \frac{\mu_0 \left|\mathbf{j}\right| R^2}{2r}, \quad A_r = A_\theta = 0, \quad A_z = -\frac{\mu_0 \left|\mathbf{j}\right| R^2}{2}\ln r . \quad r \ge R \quad (29.b)$$

In terms of either the magnetic field $\mathbf{B}$ or vector potential $\mathbf{A}$ given by (29), the energy density distributions (28) inside the current-carrying wire are identical for the fields-based and potentials-based approaches and are given by

$$u_B = u_A = \frac{\mu_0 \left|\mathbf{j}\right|^2 r^2}{8} . \qquad r \le R \quad (30.a)$$

Outside, the $\mathbf{B}$- and $\mathbf{A}$-based energy density distributions are similarly identical, and are given by

$$u_B = u_A = \frac{\mu_0 |\mathbf{j}|^2 R^4}{8r^2}. \qquad r \geq R \qquad (30.b)$$

Thus for the case of a current-carrying wire the energy density distributions $u$ for the two field variables $\mathbf{B}$ and $\mathbf{A}$, unlike the associated power flux distributions $\mathbf{S}$, are identical in their distributions, both inside and outside the wire. This is not the case for the following example where they differ substantially.

4.4 Magnetic Solenoid

As a fourth example we consider a (near-infinite-length) cylindrical magnetic solenoid of radius $R$, containing a uniform-density magnetic field $\mathbf{B}$ oriented in the $z$ direction. The $\mathbf{A}$ and $\mathbf{B}$ field components in cylindrical coordinates, related by $\mathbf{B} = \nabla \times \mathbf{A}$, are determined from $\int \mathbf{A} \cdot \mathbf{dl} = \int \mathbf{B} \cdot \mathbf{da}$,

$$B_r = B_\theta = 0, \quad B_z = B, \quad A_r = A_z = 0, \quad A_\theta = \frac{|\mathbf{B}| r}{2}, \qquad r \leq R \qquad (31.a)$$

$$B_r = B_\theta = B_z = 0, \quad A_r = A_z = 0, \quad A_\theta = \frac{|\mathbf{B}| R^2}{2r}. \qquad r \geq R \qquad (31.b)$$

In terms of the magnetic field $\mathbf{B}$, the energy densities (28) inside and outside the solenoid are given by

$$u_B = \frac{|\mathbf{B}|}{2\mu_0} \qquad r \leq R \qquad (32.a), \qquad u_B = 0 \qquad r \geq R \qquad (32.b),$$

with associated energies $E_B = \int u_B dV$ per unit length $L$ given by

$$\frac{E_B}{L} = \frac{\pi R^2 |\mathbf{B}|^2}{2\mu_0} \qquad r \leq R \qquad (33.a), \qquad \frac{E_B}{L} = 0 \qquad r \geq R \qquad (33.b).$$

That is, all the magnetostatic energy is confined *within* the solenoid.

In terms of the vector potential $\mathbf{A}$ associated with the magnetic field $\mathbf{B}$ the energy densities given by (9), $u_A = \frac{1}{2\mu_0} \sum_i |\nabla A_i|^2$, inside and outside the solenoid are given by

$$u_A^i = \frac{|\mathbf{B}|^2}{4\mu_0} \qquad (34.a), \qquad u_A^o = \frac{R^4 |\mathbf{B}|^2}{4\mu_0 r^4} \qquad (34.b),$$

where in calculating the above we again express the vector potential in its Cartesian form before calculating the derivatives, i.e.,

$$A_x = A_\theta(x,y)\sin\theta = A_\theta(x,y)\frac{y}{\sqrt{x^2+y^2}}, \quad A_y = A_\theta(x,y)\cos\theta = A_\theta(x,y)\frac{x}{\sqrt{x^2+y^2}}. \quad (35)$$

The associated energies $E_A = \int u_A dV$ per unit length $L$ are given by

$$\frac{E_A}{L} = \frac{\pi R^2 |\mathbf{B}|^2}{4\mu_0} \qquad r \le R \qquad (36.a), \qquad \frac{E_A}{L} = \frac{\pi R^2 |\mathbf{B}|^2}{4\mu_0} \qquad r \ge R \qquad (36.b),$$

i.e., are the same inside and outside the solenoid. Comparison of (36) with (33) shows that the net energies derived either in terms of the fields-based or potentials-based energy densities, though differently distributed, are equal, one being confined wholly within the solenoid $(E_B)$, the other $(E_A)$ being distributed 50% inside, 50% outside.

Given that all of the magnetic flux is confined to the interior of the solenoid, none outside, one infers (correctly) that there are no magnetic effects to be detected by classical charge motion outside, as the Lorentz force $\mathbf{F} = q\mathbf{v} \times \mathbf{B} = 0$. From a quantum viewpoint, however, despite the inability to detect classical charge effects exterior to the solenoid, at the quantum level quantum interference effects of the vector potential $\mathbf{A}$ exterior to the solenoid *can* be detected via $\mathbf{B} = \nabla \times \mathbf{A} \Rightarrow \int \mathbf{A} \cdot \mathbf{dl} = \int \mathbf{B} \cdot \mathbf{da}$ (Aharonov-Bohm, or A-B effect). To the degree that one finds it intuitively appealing to consider that registration of a quantum effect exterior to the solenoid might well be due to associated changes in the vacuum's energy density surrounding the solenoid when energized, we note that in the potentials-based approach magnetostatic energy as defined in (36) does in fact reside not only in the region interior to the solenoid, but exterior to the solenoid as well. Though not absolutely required to be the case in the classical picture, this alternative, mathematically congruent finding dovetails in a reasonable way with registration of the A-B effect.

## 5. Discussion

In the application of electromagnetic principles there has been a continuing development of various alternatives with regard to definitions involving the distributions of energy density and momentum transfer by EM fields defined in terms of the variables $(\mathbf{E}, \mathbf{B}, \mathbf{A}, \varphi)$. This is a consequence of the fact that the distributions are not uniquely determined by Maxwell's equations. Since in EM calculations all of the various (viable) options lead to identical predictions and outcomes with regard to net integrated energy density and power flux, from a mathematical viewpoint they are found to be equivalent as to net results. Therefore, strictly speaking, it is generally considered to be a matter of *aesthetic choice* as to which of the various approaches are used.[4] Nonetheless, given the vagaries of misinterpretation that can occur in application, it appears that the potentials-based approach considered herein, being one that follows from a well-defined mathematical structure implicit in the theory of partial differential equations, has much to offer and therefore comes well-recommended as a canonical procedure.

**Acknowledgements** I wish to thank M. Ibison for valuable discussion during the development of this effort.

## References

1. Feynman, R., Leighton, R., Sands, M.: The Feynman Lectures on Physics, Vol. II. Addison-Wesley Pub. Co., Menlo Park, CA, (1963), paragraph 27-5
2. Slepian, J.: J. Appl. Phys. 13, 512-518 (1942)
3. Backhaus, U., Schafer, K.: Am. J. Phys. **54**, 279-280 (1986)
4. Lai, C. S.: Am. J. Phys. **49**, 841 (1981)
5. Goldhaber, A. S., Nieto, M. M.: Rev. Mod. Phys. **43**, 277-296 (1971)
6. Zachmanoglou, E., Thoe, D.: Introduction to Partial Differential Equations with Applications, Dover, New York (1986)
7. Ribaric, M., Sustersic, L.: Conservation Laws and Open Questions of Classical Electrodynamics. World Scientific Pub. Co., Singapore (1990)
8. Jeffries, C.: SIAM Review **34**, 386-405 (1992)
9. Feynman, R., Leighton, R., Sands, M.: The Feynman Lectures on Physics, Vol. II. Addison-Wesley Pub. Co., Menlo Park, CA, (1963), paragraph 14-3

[4]As to whether the various options might be differentiated on the basis of tests by use of other than electromagnetic means, the answer is yes. Due to the difference in energy density *distributions* of the various energy density/power flux alternatives, discrimination among them could, at least in principle, be discerned on the basis of gravitational energy density measurements.